\documentclass{aastex631}

\usepackage{color}
\usepackage{hyperref}
\usepackage{epstopdf}
\usepackage{graphicx}
\usepackage[FIGTOPCAP]{subfigure}
\usepackage{CJKutf8}
\usepackage{amsmath}
\usepackage{longtable}
\usepackage{threeparttable}
\usepackage{booktabs}
\usepackage{comment}

\begin{document}
\title{Neutron Star Merger Rates from Multi-messenger Observations: Clues to the Physical Origin of the Short and Long-short Gamma-ray Bursts}

\author[0000-0003-4977-9724]{Zhi-Ping Jin}
\affiliation{Purple Mountain Observatory, Chinese Academy of Sciences, Nanjing 210023, China}
\affiliation{School of Astronomy and Space Science, University of Science and Technology of China, Hefei 230026, China}
%\email{jin@pmo.ac.cn}

\author[0000-0001-9626-9319]{Yuan-Zhu Wang}
\affiliation{Institute for Theoretical Physics and Cosmology, Zhejiang University of Technology, Hangzhou, 310032, People's Republic of China}
%\email{vamdrew@zjut.edu.cn}

\author[0000-0001-5087-9613]{Yin-Jie Li}
\affiliation{Purple Mountain Observatory, Chinese Academy of Sciences, Nanjing 210023, China}
%\email{liyinjie@pmo.ac.cn}

\author[0000-0002-8385-7848]{Yun Wang}
\affiliation{Purple Mountain Observatory, Chinese Academy of Sciences, Nanjing 210023, China}
%\email{wangyun@pmo.ac.cn}

\author[0000-0002-0556-1857]{Hao Wang}
\affiliation{Purple Mountain Observatory, Chinese Academy of Sciences, Nanjing 210023, China}
%\email{haowang@pmo.ac.cn}

\author[0000-0001-9120-7733]{Shao-Peng Tang}
\affiliation{Purple Mountain Observatory, Chinese Academy of Sciences, Nanjing 210023, China}
%\email{tangsp@pmo.ac.cn}

\author[0000-0002-9758-5476]{Da-Ming Wei}
\affiliation{Purple Mountain Observatory, Chinese Academy of Sciences, Nanjing 210023, China}
\affiliation{School of Astronomy and Space Science, University of Science and Technology of China, Hefei 230026, China}
%\email{dmwei@pmo.ac.cn}

\correspondingauthor{Zhi-Ping Jin, Yuan-Zhu Wang}
\email{jin@pmo.ac.cn, vamdrew@zjut.edu.cn}

\begin{abstract}
Short and long-short gamma-ray bursts (GRBs) are widely believed to be powered by neutron star mergers. In this work, we calculate local rate of such GRBs and find a relatively high value of $\sim 786-2468~{\rm Gpc^{-3}~yr^{-1}}$ when including the very narrow collimation event GRB 061201. Considering that its redshift is not very reliable, after excluding this event, the rate is $\sim 195-666~{\rm Gpc^{-3}~yr^{-1}}$. We also calculate the electromagnetically (EM) bright neutron star merger rate inferred from the LIGO/Virgo/KAGRA observations up to the end of the first epoch of the O4 run, and derive a rate of $\sim 66-347~{\rm Gpc^{-3}~yr^{-1}}$. This rate is somewhat lower than the value obtained from the GRBs, even after excluding GRB 061201. The non-detection of any viable EM bright merger in the O4b and O4c observing runs favors an even lower rate, which starts to challenge the neutron star merger origin of the short and long-short GRBs and may suggest additional contribution from the mergers of other compact object (like the neutron star-white dwarf) binaries, as speculated initially by \citet{2007MNRAS.374L..34K}  in interpreting the long-short event GRB 060614. 
\end{abstract}
\keywords{Gamma-ray bursts (629); Neutron stars (1108); Gravitational waves (678)}

\section{Introduction}
\label{Sec:intro}
Neutron star mergers are events involving the coalescence of at least one neutron star with another compact object, including binary neutron star(BNS) mergers and neutron star–black hole(NSBH) mergers. These events can produce phenomena such as gamma-ray burst(GRB), gravitational wave(GW)\citep{1989Natur.340..126E} and kilonova(KN)\citep{1998ApJ...507L..59L}, and represent one of the primary astrophysical sources of heavy elements heavier than iron in our universe\citep{1974ApJ...192L.145L}. Various methods were employed to estimate the rate of neutron star mergers, including observations of GW events, short and long-short GRBs, KNe, and Galactic pulsar binaries \citep[see a recent review in][]{2022LRR....25....1M}. 

It has long been speculated that short GRBs (SGRBs) may be associated with neutron star mergers. In 2005, the discovery of X-ray and optical afterglows from SGRBs enabled precise localization of these events. Distances were determined either by measuring the redshift of the host galaxies or directly from the afterglows\citep{2005Natur.437..851G,2005Natur.437..845F}. Additional evidence supporting the neutron star merger origin came from the types of host galaxies, the locations within those galaxies, and the absence of supernova signatures in late-time afterglow observations. The connection between SGRB and KN, established through GRB 130603B in 2013, further strengthened the merger origin scenario\citep{2013Natur.500..547T}. Surprisingly, in 2015, a re-analysis of archival data by \cite{2015NatCo...6.7323Y} revealed a KN signature in the afterglow of the long-duration GRB 060614, providing further evidence in support of the long-standing conjecture that this particular long GRB actually originated from a compact binary merger. Subsequently, KN signatures have been detected in additional long-duration GRBs, confirming their origin from compact binary mergers. These events are also known as long-short GRBs (lsGRBs). SGRBs have been widely discussed in the literature as probes for estimating the neutron star merger rate, though several challenges remain, such as the viewing angle and sensitivity of detectors, as well as the highly beamed nature of SGRBs. In \cite{2018ApJ...857..128J}, we used a sample of SGRBs and lsGRBs detected by the \textit{Swift} BAT instrument, which ensured a uniform viewing angle and the sensitivity. By considering only SGRBs with redshifts lower than 0.2 and opening angle measurements available (to minimize selection effects and ensure homogeneity), we derived a lower limit on the neutron star merger rate in the local universe based on both SGRBs and lsGRBs. Since then, more SGRBs and lsGRBs meeting these criteria have been detected \citep{2018NatCo...9.4089T, 2019MNRAS.489.2104T, 2019ApJ...883...48L, 2022Natur.612..228T, 2022Natur.612..223R, 2022Natur.612..232Y, 2024Natur.626..742Y, 2024Natur.626..737L}, 
in principle, they were expected to provide a stronger constraints on the neutron star merger rate.

The first detection of binary neutron star merger via gravitational waves was GW170817 \citep{2017PhRvL.119p1101A}. It was also accompanied by the detection of a short GRB 170817A and a KN  \citep{2017ApJ...848L..12A}. Following the discovery of GW170817, it was used to estimate the merger rate of binary neutron stars \citep{2017PhRvL.119p1101A}. Recently, the LIGO Virgo KAGRA collaboration (LVKC) released the Gravitational-Wave Transient Catalog 4.0 (GWTC-4.0). However, due to various observational limitations, no electromagnetic counterparts have been identified for any subsequent NS merger events. Despite the absence of electromagnetic observations, the merger rates for both BNS and NSBH systems can be estimated by accounting for their false alarm rates and detection sensitivities \citep{2025arXiv250818083T, 2025arXiv250708778A}.

In this work, we refine the neutron star merger rate based on the latest observational data of SGRBs and lsGRBs, as detailed in Section \ref{Sec:SGRBrate}. In Section \ref{Sec:GWrate}, we compare this result with the electromagnetically (EM) bright neutron star merger rate we derived from gravitational wave observations and identify a noticeable inconsistency between them. Finally, we provide a detailed discussion on the possible causes of this discrepancy in section \ref{Sec:Discussion}.

\section{Neutron Star Merger Rate from SGRBs and lsGRBs}
\label{Sec:SGRBrate}

Once a given instrument detects an individual GRB originating from a nearby merger, its contribution to the local neutron star merger rate can be expressed as:

\begin{equation}
R_\text{nsm} = \frac{4\pi}{\left(1 - \cos\left(\max[\theta_\text{j}, \theta_\text{v}]\right)\right) \text{FoV}~T~V_{\text{c(z=0.2)}}}.
\end{equation}
Here, $\theta_\text{j}$ and $\theta_\text{v}$ are the half opening jet angle and the viewing angle of the GRB, respectively. $\text{FoV}$ is the field of view of this instrument, for \textit{Swift} BAT, its about 1.4-2.4 steradians. $T$ represents the total observation time during the instrument's lifetime, $V_{\text{c(z=0.2)}}$ is the total comoving volume for redshifts less than 0.2. 

The total neutron star merger rate inferred from GRB observations is obtained by summing up the contributions from all GRBs detected by a specific instrument. For merger rates derived from GRBs detected by different instruments cannot be directly summed. Instead, the merger rate must be estimated separately for each instrument. Due to limitations in follow-up observations, some events were missing from the statistical sample. As a result, the neutron star merger rate obtained through this method should be regarded only as a lower limit. This issue is particularly severe for \textit{Fermi} GBM, which has a large localization error hence many detected GRBs do not have their distances measurement. In this work, we calculate the contribution only from GRBs detected by the \textit{Swift} BAT instrument. 

For the BAT on board \textit{Swift}, its FoV and sensitivity change as a function of the partial coding fraction, which depends on the burst’s incident angle \citep{2005SSRv..120..143B}. In this work, we estimate BAT's sensitivity following the approaches introduced in \citet{2018ApJ...857..128J}. \cite{2016ApJ...829....7L} have shown that there is a relation between the $T_{90}$ of bursts and the minimum observable time-average energy flux in the $15-150$ keV band: the minimum flux is $3\times10^8 {\rm \ erg\ cm^{-2} \ s^{-1} }$ when $T_{90}=1{\rm \ s}$. The minimum flux corresponding to other given duration can be derived with $\mathcal{F}_{\rm min} = 3\times10^8/\sqrt{T_{90}} {\rm \ erg\ cm^{-2} \ s^{-1} }$. A burst with flux $\mathcal{F}_{\rm min}$ can be detected only when it's incident angle is small, i.e., it's emission is fully coded by BAT. If a burst’s incident angle is large, and the detector’s plane is partially coded, the effect of partial coding fraction $P_{\rm f}$ can be expressed with the ``effective on-axis exposure time" $T_{\rm eff} = P_{\rm f} \times T_{90}$. Considering the above relations, we calculate the smallest partial coding fraction at any given distance with:

\begin{equation}
\label{eq:pfmin}
    P_{\rm f,min} = \left[ \frac{3\times10^8 {\rm \ erg\ cm^{-2} \ s^{-1} }}{(\frac{D_{\rm L0}}{D_{\rm L}})^2(\frac{1+z}{1+z_0})\mathcal{F}_0} \right]^2/T_{90},
\end{equation}
where $z_0$, $D_{\rm L}$, and $\mathcal{F}_0$ are the observed redshift, luminosity distance, and time-averaged flux during $T_{100}$ respectively, of a given burst. 
The resulting $P_{\rm f,min}$ determines the corresponding BAT FoV. The relation ${\rm FoV}(P_{\rm f,min})$, was presented in \cite{2005SSRv..120..143B}, and here we use their simulated curve adjusted for off-axis projection effects. 
The spacetime volume covered by searching for a given burst is then calculated by: 

\begin{equation}
\label{eq:VT}
    \left \langle VT \right \rangle = 0.78\ T_{\rm obs} \int_{0}^{0.2} \frac{{\rm FoV}(P_{\rm f,min})}{4\pi}\frac{1}{1+z}\frac{{\rm d}V_{\rm c}}{{\rm d}z}{\rm dz},
\end{equation}
where the factor 0.78 represents the fraction of time that BAT spends on searching for GRBs according to \cite{2016ApJ...829....7L}. 

Assuming a negligible evolution of rate in the local universe, we first calculate the apparent event rate (without opening angle correction), $\mathcal{R_{\rm app}}$ , in the local ($z<0.2$) universe under the frame-work of Baysian inference. The posterior distribution of $\mathcal{R_{\rm app}}$ can be written as:

\begin{equation}
    {\rm P}(\mathcal{R_{\rm app}}) \propto \mathcal{ L}_{\rm Poisson}(1|\Lambda)\times{\rm P}'(\mathcal{R_{\rm app}})
\end{equation}
where $\mathcal{ L}_{\rm Poission}(1|\Lambda)$ is the likelihood of observing one event from a Poisson distribution with a mean number $\Lambda = \mathcal{R_{\rm app}}\left \langle VT \right \rangle$, and ${\rm P}'(\mathcal{R_{\rm app}})$ is the prior. By adopting a uniform prior for the apparent event rate, we derive the constraints on  $\mathcal{R_{\rm app}}$ for GRB 060614, GRB 061201, GRB 160821B, and GRB 211211A, respectively. The results are listed in Tabel \ref{tab:grb_prop}. Similar to the results presented in \cite{2018ApJ...857..128J}, the apparent rates for these events are close to each other, suggesting that these GRBs are bright and can be detected within the FoV $\lesssim 2.4$ sr at $z \sim 0.2$. 

\begin{figure*}
\label{fig:rates}
	\centering  
\includegraphics[width=0.48\linewidth]{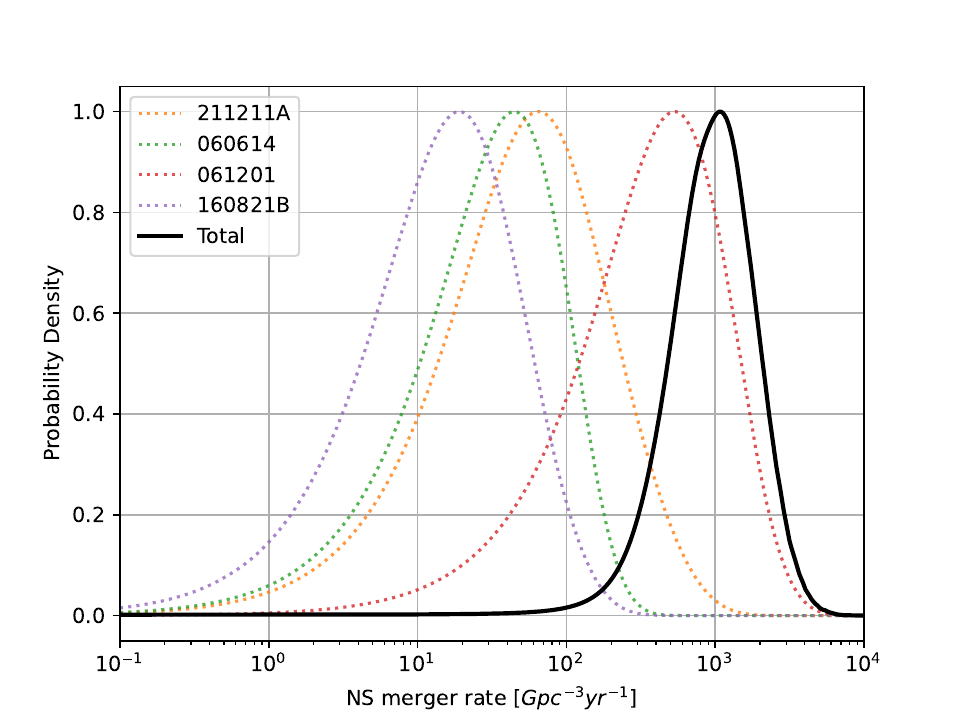}
\includegraphics[width=0.48\linewidth]{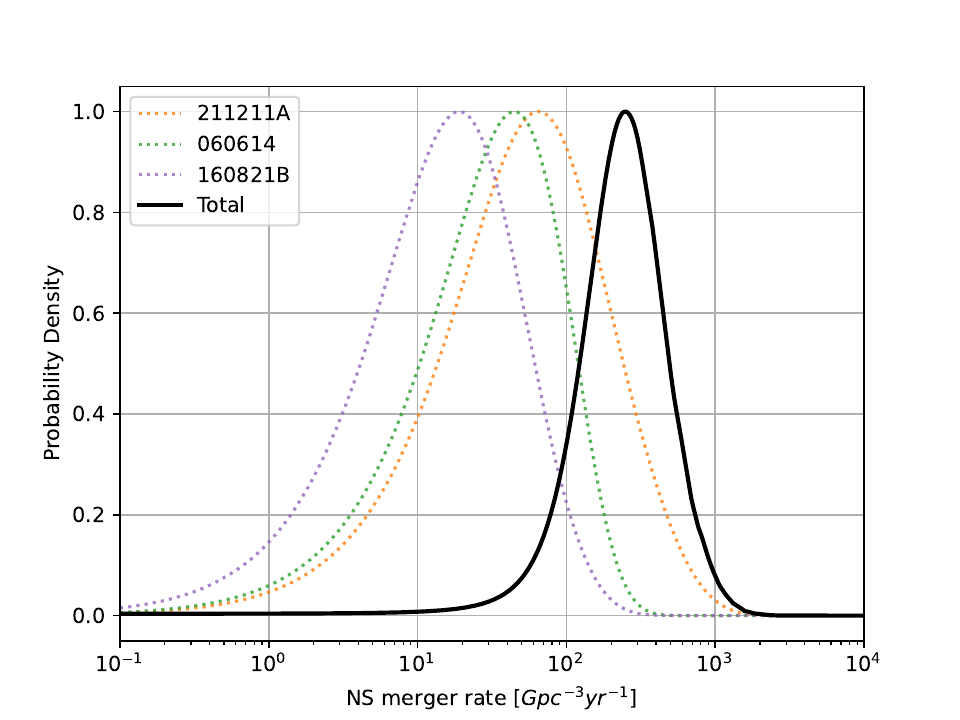}
\caption{Neutron star merger rates derived from short and long-short GRBs. Left: results include GRB 061201. Right: result exclude GRB 061201.}
\end{figure*}

Next, we calculate the local neutron star merger rates $\mathcal{R}_{\rm NS}$ by performing geometry corrections on the search volumes of each events. Following \cite{2018ApJ...857..128J}, the posterior probability density of $\mathcal{R}_{\rm NS}$, marginalized over the half-opening angles $\theta_{\rm j}$, can be expressed as:

\begin{equation}
    {\rm P}(\mathcal{R_{\rm NS}}) = \int_{\theta_{\rm j,min}}^{\theta_{\rm j,max}} \mathcal{ L}_{\rm Poisson}(1|\Lambda')\ {\rm P}'(\mathcal{R_{\rm app}})\ {\rm P}'(\theta_{\rm j})\ {\rm d}\theta_{\rm j}
\end{equation}
in which $\Lambda' = \mathcal{R_{\rm NS}}\frac{\theta_{\rm j}^2}{2}\left \langle VT \right \rangle$. As discussed in \cite{2018ApJ...857..128J}, the measurement of redshift for GRB061201 is not secure. Therefore, we derive rates  with and without this event, the results are $1425.66^{+1041.85}_{-639.56}$ and $354.84^{+311.61}_{-160.14}$  respectively. More details are shown in Table \ref{tab:grb_prop} and Figure \ref{fig:rates}.

Comparing with \cite{2018ApJ...857..128J}, the values of $\mathcal{R_{\rm NS}}$ for each event obtained in this work are smaller. This is primarily because each event is considered as an unique sub-class of GRB hence the rate will decrease with the total observational time. At the same time, the number of events is increasing, leading the total merger rate to approach the true time-averaged rate.

\section{Comparison with the GW Derived Merger Rate}
\label{Sec:GWrate}
The LVKC has recently released the GWTC-4.0, providing updated estimates of the local merger rates for BNS and NSBH systems \citep{2025arXiv250818083T}. The reported rates are \(7.6\text{--}250  \text{Gpc}^{-3} \text{yr}^{-1}\) for BNS and \(9.1\text{--}84  \text{Gpc}^{-3} \text{yr}^{-1}\) for NSBH mergers. 
For the BNS population, both the primary and secondary masses are $\lesssim 2 M_\odot$, it is expected that they can all produce electromagnetic counterparts, such as GRBs and KNe. However, for the NSBH population, where the primary mass is $M_{1} \sim 9 M_\odot$ and the secondary mass is $M_{2} \lesssim 2 M_\odot$, a mass ratio which implies that most NSBH merger events cannot produce an electromagnetic counterpart. 
Therefore, the BNS merger rate is fundamentally representative of the event rate for merger origin GRBs. 

In this work, we derive the local EM bright neutron star merger rate from gravitational wave observations as follows. First, we select BNS or NSBH events with potential to produce GRBs. We found that GW170817, GW190425, GW200115, and GW230529 satisfy such criteria, with the HasRemnant parameter is greater than 0, see Table \ref{tab:gw_prop}. Then,  we perform hierarchical Bayesian inference and derive the local merger rate (together with the hyper-parameters of mass function) for the EM bright mergers \citep{2021ApJ...923...97L,2023PhRvX..13a1048A}. We apply a simple mass function for the BNS and NSBH mergers 
\begin{equation}
\begin{aligned}
    P(m_1,m_2 &|\alpha,\beta,m_{\rm min},m_{\rm max,NS},m_{\rm max,BH})\propto \\ &\mathcal{PL}(m_1|\alpha,m_{\rm min},m_{\rm max,BH})\times \\
    &\mathcal{PL}(m_2|\alpha,m_{\rm min},m_{\rm max,NS})\times (m_2/m_1)^\beta,
\end{aligned}
\end{equation}
where $\mathcal{PL}$ is the Truncated PowerLaw, $m_{\rm min}$ is the lower cutoff, $m_{\rm max,NS}$ and $m_{\rm max,BH}$ are the upper cutoff for the NS and BH, $\alpha$ and $\beta$ are the PowerLaw indexes for the mass function and pairing function. We assume the mergers are uniformly distributed in the source frame. 
Then the EM bright merger rate is inferred as $164.51^{+182.59}_{-98.16}\text{Gpc}^{-3} \text{yr}^{-1}$, slightly higher than the value $13-170\text{Gpc}^{-3} \text{yr}^{-1}$ derived by the SIMPLE UNIFORM BNS model \citep{2025arXiv250818083T}.
The upper bound of BNS merger rate is consistent with the lower bound of our total rate derived from GRBs after excluding GRB 061201, i.e. $354.84^{+311.61}_{-160.14}\text{Gpc}^{-3} \text{yr}^{-1}$. This suggests that GRB 061201 likely originated from a higher redshift, a possibility as discussed in \cite{2007A&A...474..827S}. 

\begin{figure*}
	\centering  
\includegraphics[width=0.48\linewidth]{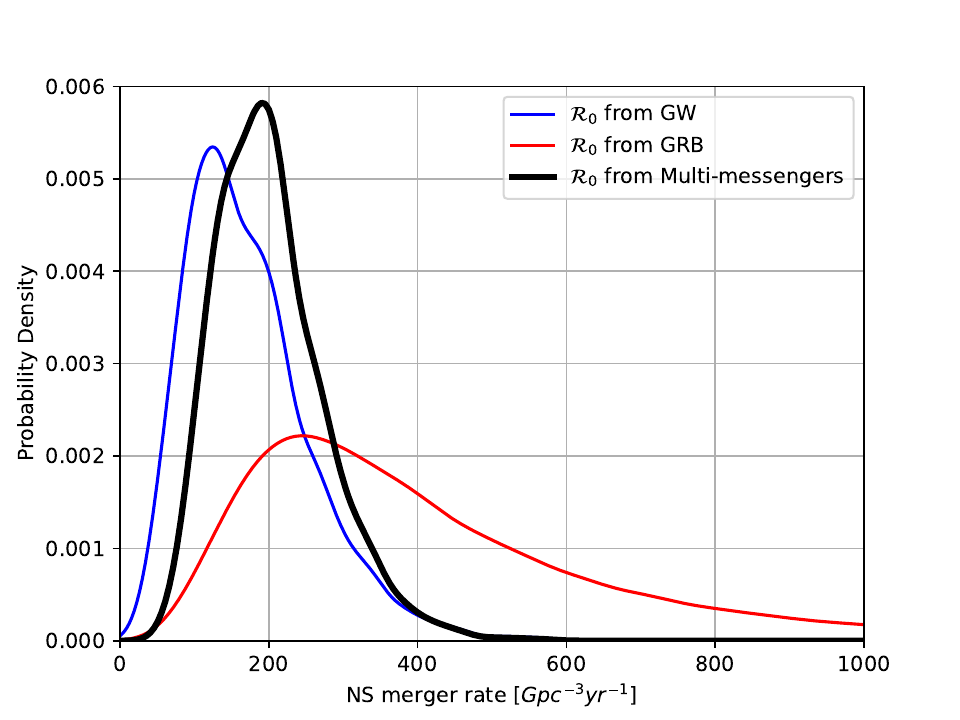}
\caption{Probability density function for the local EM bright merger rate derived from different approaches. The blue curve is the rate obtained by performing hierarchical Bayesian inference on the chosen GW events; the red curve is the rate calculated from GRBs as introduced in Section \ref{Sec:SGRBrate}; the black solid curve is the multi-messenger result derived from both GW and GRB observations.}
\label{fig:rates_mma}
\end{figure*}

Since GWs and GRBs are observed by distinct instruments, the joint constraint can be easily calculated by $P_{\rm MM}(\mathcal{R}_{\rm NS}) \propto P_{\rm GW}(\mathcal{R}_{\rm NS})\times P_{\rm GRB}(\mathcal{R}_{\rm NS})$, where $P_{\rm MM}$, $P_{\rm GW}$, and $P_{\rm GRB}$ are the posterior probability density function for the multi-messengers approach, the GW approach, and GRB approach respectively. Finally, the multi-messengers approach yields a local merger rate of $193.28^{+80.62}_{-63.95}\text{Gpc}^{-3} \text{yr}^{-1}$. We present these results in Figure \ref{fig:rates_mma}.

It is noteworthy that our calculation depends on the GRB jet opening angle to correct the detection rate. GRB 080905A had too sparse observational data to determine its jet opening angle. GRB 100216A and GRB 111005A lacked afterglow detections, and their redshifts were determined solely from their prospective host galaxies. These bursts were not included in our event rate calculation. Consequently, our result represents a lower limit on the event rate. If these GRBs were included under the assumption that their jet opening angles are comparable to those of other SGRBs, the derived event rate would be further increased. 

To summarize, we find a relatively high value of the event rate for the short and long-short GRBs $354.84^{+311.61}_{-160.14}\text{Gpc}^{-3} \text{yr}^{-1}$, even after excluding the very narrowly collimated event GRB 061201 and not well observed events GRB 080905A, GRB 100216A and GRB 111005A. 
Although still within uncertainties, our inferred rate is higher than the binary neutron star merger rate of \(7.6\text{--}250  \text{Gpc}^{-3} \text{yr}^{-1}\) inferred with GWTC-4.0 catalog\citep{2025arXiv250818083T}. The non-report of any viable binary neutron star merger rate in the O4b and O4c runs is in favor an even lower ${\cal R_{\rm BNS}}$\citep{2025arXiv250708778A}, which starts to challenge the binary neutron star merger origin of the short and long-short GRBs and may suggest additional contribution from the mergers of other compact object (like the neutron star-white dwarf) binaries, as speculated initially by \citet{2007MNRAS.374L..34K} in interpreting the long-short event GRB 060614. 

\section{Discussion and Conclusion}
\label{Sec:Discussion}
In this work, we only considered GRBs observed by \textit{Swift} BAT. As can be seen from Table \ref{tab:grb_prop}, Fermi GBM was also triggered by 6 nearby bursts in 17 years, which is only slightly fewer than the 8 (GRB 150101B and GRB 230307A are not included) that triggered \textit{Swift} BAT in 20 years. However, Fermi GBM has a FoV covering a full 4$\pi$ steradians, meaning it observes about 65\% of the sky, apart from Earth occulting. Therefore, even though the detection rate of Fermi GRM is comparable to that of \textit{Swift} BAT, the event rate inferred from its data is considerably lower. This is due to Fermi GBM’s lower localization accuracy and the absence of onboard follow-up instruments such as \textit{Swift}’s XRT and UVOT, the fraction of Fermi GBM-triggered GRBs confirmed as nearby bursts is relatively low. In fact, among the 6 nearby GRBs triggered by Fermi, aside from GRB 170817A (associated with the gravitational wave event GW 170817) and the very bright GRB 230307A (which also triggered numerous other instruments, with its position determined by them), the other 4 were all quickly localized by Swift. During its 7.5-year lifetime, HETE-2 FREGATE (with a FOV of about 3 steradians) detected only one nearby GRB. Therefore, the event rate derived from HETE-2 would be very low, unless GRB 050709 possessed a jet opening angle as small as that of GRB 161201.

In our previous work \citep{2018ApJ...857..128J}, we estimated the local neutron star merger rate density to be $1109^{+1432}_{-657}~\text{Gpc}^{-3}~\text{yr}^{-1}$, or $162^{+140}_{-83}~\text{Gpc}^{-3}~\text{yr}^{-1}$ if the narrowly beamed GRB 061201 is excluded. Eight years later, our updated estimates are $1425.66^{+1041.85}_{-639.56}~\text{Gpc}^{-3}~\text{yr}^{-1}$ when GRB 061201 is included, and $354.84^{+311.61}_{-160.14}~\text{Gpc}^{-3}~\text{yr}^{-1}$ when it is excluded. This is mainly attributed to the addition of the new sample, GRB 211211A, to the dataset, which also has a relatively narrow jet opening angle. 
At the meanwhile, BNS merger rate has decreased by a factor of $\sim~2$ during the end of LIGO O3 to O4a \citep{2025arXiv250818083T}. 
Although the findings are based on only a handful of events for both GRBs and GW events, a preliminary tension has emerged and must be tested by further observations. It is interesting because if the neutron star merger rate inferred from short and long-short GRBs is larger than that inferred from GW events, it means some of short and long-short GRBs may not comes from neutron star merger, even though the sample GRB 060614\citep{2015NatCo...6.7323Y}, GRB 160821B\citep{2018ApJ...857..128J, 2019MNRAS.489.2104T, 2019ApJ...883...48L} and GRB 211211A\citep{2022MNRAS.510.1902T} are all associated with kilonovae candidates. This does not even include the long-short burst GRB 060505, which could also has kilonovae signal and be merger origin. This implies that a fraction of short and long-short GRBs associated with kilonovae may not come from neutron star mergers.

Another possible explanation is jet wobbling or precession, which can alter the jet structure from a collimated cone to a ring. This phenomenon has been seen in numerical simulations of both collapsars and neutron star–black hole mergers \citep{2022ApJ...933L...9G,2023ApJ...954L..21G,2023PhRvD.107l3001H}, where a ring structure is particularly evident in the latter case. Such a ring-shaped structure may also help explain the shallow post–jet break decay observed in some GRBs (Wang et al., in preparation). In this situation, the solid angle of a ring is approximately $4\pi\theta_r\theta_c$, where $\theta_r$ and $\theta_c$ are the latitude and thickness of the ring, respectively. If we assume that a fraction of BNS mergers also produce ring-shaped jets, their associated short GRBs could be visible over a solid angle much larger than $\pi\theta_c^2$. Therefore, the BNS merger rate inferred from short GRBs may be overestimated. This scenario also worth further investigation through numerical simulations and observations. 

\section*{Acknowledgments}
We thank Prof. Yi-Zhong Fan for the stimulating discussion. 
We acknowledge the use of the public data from the Swift data archive. 
We also acknowledge the use of public data from the LIGO and Virgo observatories. 
This work is supported by the Natural Science Foundation of China (grant Nos. 12225305, 12233011, 12203101, 12503059, 12473049, and 12321003), the National Key R\&D Program of China (grant Nos. 2024YFA1611704 and 2024YFA1611700), the Strategic Priority Research Program of the Chinese Academy of Sciences (grant No. XDB0550400). 
Y.J.L. is supported by the General Fund (No. 2024M753495) of the China Postdoctoral Science Foundation. 
Y.W. is supported by the Jiangsu Funding Program for Excellent Postdoctoral Talent (grant No. 2024ZB110), the Postdoctoral Fellowship Program (grant No. GZC20241916) and the General Fund (grant No. 2024M763531) of the China Postdoctoral Science Foundation. 

\vspace{5mm}
\facilities{Swift(BAT), Fermi(GBM), HETE-2, LIGO}

\begin{table*}[!htbp]
\flushleft
\caption{The properties of nearby (redshift $z < 0.2$) SGRBs and lsGRBs, including their jet break time ($t_{\mathrm{j}}$, in days), jet opening angle ($\theta_{\mathrm{j}}$, in radians), the average flux ($\mathcal{F}_{100}$, in $\mathrm{erg~cm^{-2}~s^{-1}}$), the derived apparent rate $\mathcal{R_{\rm app}}$ and event rate $\mathcal{R_{\rm NS}}$(in $\text{Gpc}^{-3}~\text{yr}^{-1}$, the values represent the median, with a 68\% confidence interval). The flux data are obtained from the \textit{Swift/BAT Gamma-Ray Burst Catalog} \url{https://swift.gsfc.nasa.gov/results/batgrbcat/}.}
\label{tab:grb_prop}
\begin{center}
\begin{tabular}{lcccccccc}
\toprule
Facility & GRB & z & $t_{\text{j}}$ & $\theta_{\text{j}}$ & $\mathcal{F}_{100}$ & $\mathcal{R_{\rm app}}$ & $\mathcal{R_{\rm NS}}$ & Reference\\
\hline
HETE-2 & 050709 & 0.160 & $<$1.4 & $<$0.14 & - & -   & - &(1)(2)\\
Swift & 060505 & 0.089 & - & - & 1.13E-7 & - & - & (3)\\
Swift & 060614 & 0.125 & 1.4 & 0.08-0.09 & 1.05E-7 & $0.27^{+0.26}_{-0.16}$ & $75.5^{+72.72}_{-43.55}$ & (4)\\
Swift & 061201 & 0.111 & 0.03 & 0.02-0.03 & 4.00E-7 & $0.28^{+0.27}_{-0.16}$ & $964.49^{+1002.05}_{-567.71}$  & (5)\\
Swift & 080905A & 0.1218 & - & - & 1.29E-7 & - & - & (6)\\
Swift/Fermi & 100216A & 0.0378 & - & - & 9.69E-8 & - & - & (7)\\
Swift & 111005A & 0.01326 & - & - & 2.05E-8 & - & - & (8)\\
Swift/Fermi & 150101B & 0.1341 & 10-15 & $\theta_{\text{v}}$=0.23 & 1.52E-6$^{\rm a}$ & - & - & (9)\\
Swift/Fermi & 160821B & 0.1613 & 0.7 & 0.1 & 2.15E-7  & $0.29^{+0.28}_{-0.16}$ & $43.6^{+57.14}_{-27.15}$  & (10-12)\\
Fermi & 170817A & 0.0098 & 150-160 & $\theta_{\text{v}}$=0.24-0.5 & - & - & - & (13-16)\\
Fermi/Swift & 211211A & 0.0762 & 0.463 & 0.033-0.1$^{\rm b}$ & 1.21E-6 & $0.27^{+0.26}_{-0.16}$ & $196.3^{+306.84}_{-129.84}$ & (17-19)\\
Fermi/Swift$^{\rm c}$ & 230307A & 0.0646 & 0.3/0.949 & 0.04$^{\rm b}$/0.23 & - & - & - & (20)/(21)\\
\hline
Total & - & - & - & - & - & - & $1425.66^{+1041.85}_{-639.56}$ & -\\
%${\rm Total}^{\rm d}$ & - & - & - & - & - & - & $368.56^{+310.39}_{-159.59}$ & - \\
${\rm Total}^{\rm d}$ & - & - & - & - & - & - & $354.84^{+311.61}_{-160.14}$ & - \\

\bottomrule
\end{tabular}
\end{center}

\begin{tablenotes}
\footnotesize
\item \textbf{References:}
(1)\cite{2005Natur.437..845F};  (2)\cite{2016NatCo...712898J}; 
(3)\cite{2007ApJ...662.1129O}; 
(4)\cite{2006Natur.444.1050D}; 
(5)\cite{2007A&A...474..827S}; 
(6)\cite{2010MNRAS.408..383R}; 
(7)\cite{2010GCN.10429....1P}; 
(8)\cite{2011GCN.12414....1L}; 
(9)\cite{2018NatCo...9.4089T}; 
(10)\cite{2018ApJ...857..128J};  (11)\cite{2019MNRAS.489.2104T};  (12)\cite{2019ApJ...883...48L}; 
(13)\cite{2019ApJ...870L..15L};  (14)\cite{2021ApJ...922..154M};  (15)\cite{2022MNRAS.510.1902T}; 
(16)\cite{2023ApJ...943...13W}; 
(17)\cite{2022Natur.612..228T}; (18)\cite{2022Natur.612..223R}; (19)\cite{2022Natur.612..232Y}; 
(20)\cite{2024Natur.626..742Y}; 
(21)\cite{2024Natur.626..737L}. 
\item[]$^{\rm a}$The flux is recalculated in this work. Since this is a sub-threshold detection for {\it Swift} BAT, it cannot be included in our sample statistics according to the criteria of the methodology.
\item[]$^{\rm b}$Referring to the core opening angle of the structured jet ($\theta_{\rm c}$). 
\item[]$^{\rm c}$GRB 230307A was detected by \textit{Swift} BAT but was outside the coded FoV and therefore is not included in our event rate calculation samples. 
%\item[d]Total rate after excluding GRB 061201.
\item[]$^{\rm d}$Total rate after excluding GRB 061201. 
\end{tablenotes}
\end{table*}

\begin{table*}[!htbp]
\label{tab:gw_prop}
\caption{The properties of gravitational wave events with secondary mass $<3 M_\odot$. The data are taken from the \textit{Gravitational Wave Open Science Center} \url{https://gwosc.org/eventapi/html/allevents/}, and supplemented by \cite{2017ApJ...848L..12A, 2023GCN.34148....1L, 2020GCN.26807....1L}. }

\begin{center}
\begin{tabular}{lcccccccc}
\toprule
\textbf{Event Name} & \textbf{Mass 1} & \textbf{Mass 2} & \textbf{Total Mass} & \textbf{Chirp Mass} & \textbf{Distance} & \textbf{SNR} & \textbf{HasRemnant} \\
& $[M_\odot]$ & $[M_\odot]$ & $[M_\odot]$ & $[M_\odot]$ & [Mpc] & & \\
\midrule
GW170817 & $1.46_{-0.10}^{+0.12}$ & $1.27_{-0.09}^{+0.09}$ & $2.82_{-0.09}^{+0.47}$ & $1.186_{-0.001}^{+0.001}$ & $40_{-15}^{+7}$ & 33.0 & 1 \\
GW190425 & $2.1_{-0.4}^{+0.5}$ & $1.3_{-0.2}^{+0.3}$ & $3.4_{-0.1}^{+0.3}$ & $1.44_{-0.02}^{+0.02}$ & $150_{-60}^{+80}$ & $12.4_{-0.4}^{+0.4}$ & 1 \\
GW190814 & $23.3_{-1.4}^{+1.4}$ & $2.6_{-0.1}^{+0.1}$ & $25.9_{-1.3}^{+1.3}$ & $6.11_{-0.05}^{+0.06}$ & $230_{-50}^{+40}$ & $25.3_{-0.2}^{+0.1}$ & 0 \\
GW190917\_114630 & $9.7_{-3.9}^{+3.4}$ & $2.1_{-0.4}^{+1.1}$ & $11.8_{-2.8}^{+3.0}$ & $3.7_{-0.2}^{+0.2}$ & $720_{-310}^{+300}$ & $8.3_{-0.8}^{+0.5}$ & 0 \\
GW191219\_163120 & $31.1_{-2.8}^{+2.2}$ & $1.17_{-0.06}^{+0.07}$ & $32.3_{-2.7}^{+2.2}$ & $4.31_{-0.17}^{+0.12}$ & $550_{-160}^{+240}$ & $9.1_{-0.8}^{+0.5}$ & 0 \\
GW200115\_042309 & $5.9_{-2.5}^{+2.0}$ & $1.44_{-0.28}^{+0.85}$ & $7.4_{-1.7}^{+1.7}$ & $2.43_{-0.07}^{+0.05}$ & $290_{-100}^{+150}$ & $11.3_{-0.5}^{+0.3}$ & 1 \\
GW200210\_092254 & $24.1_{-4.6}^{+7.5}$ & $2.83_{-0.42}^{+0.47}$ & $27.0_{-4.3}^{+7.1}$ & $6.56_{-0.4}^{+0.38}$ & $940_{-340}^{+430}$ & $8.4_{-0.7}^{+0.5}$ & 0 \\
GW230518\_125908 & $8.17^{+0.84}_{-0.92}$ & $1.45^{+0.13}_{-0.10}$ & $9.61^{+0.76}_{-0.79}$ & $2.80^{+0.06}_{-0.06}$ & $240^{+110}_{-100}$ & $14.2^{+0.2}_{-0.4}$ & 0 \\
GW230529\_181500 & $3.66^{+0.82}_{-1.21}$ & $1.42^{+0.60}_{-0.22}$ & $5.08^{+0.61}_{-0.60}$ & $1.94^{+0.04}_{-0.04}$ & $200^{+100}_{-100}$ & $11.6^{+0.3}_{-0.4}$ & 0.07 \\
%Events With Remnant & & & & & & & & $120^{+155}_{-78}$ \\
%Events With Remnant & & & & & & & & $164.51^{+182.59}_{-98.16}$ \\
\bottomrule
\end{tabular}
\end{center}

\end{table*}

%\newpage
\bibliography{bibliography.bib}{}
\bibliographystyle{aasjournal}

\end{document}